\documentclass[nofootinbib, superscriptaddress, twocolumn, pra]{revtex4}

\usepackage{amsmath}
\usepackage{amsthm}
\usepackage{amssymb}
\usepackage{graphicx}
\usepackage{color}
\usepackage{qcircuit}
\usepackage{braket}
\usepackage{bbm}
\DeclareMathOperator{\id}{\mathbbm{1}}

\def\ket#1{\left| #1 \right\rangle}
\def\bra#1{\left\langle #1 \right|}

\newcommand{\bx}{\mathbf{x}}

\newcommand{\beq}{\begin{equation}}
\newcommand{\eeq}{\end{equation}}
\definecolor{JM}{RGB}{4,116,149}

\begin{document}
\title{Quantum algorithm for non-homogeneous linear partial differential equations}
\author{Juan Miguel Arrazola}
\email{juanmiguel@xanadu.ai}
\author{Timjan Kalajdzievski}
\author{Christian Weedbrook}
\affiliation{Xanadu, 372 Richmond Street W, Toronto, Ontario M5V 1X6, Canada}
\author{Seth Lloyd}
\affiliation{Massachusetts Institute of Technology, Department of Mechanical Engineering,
77 Massachusetts Avenue, Cambridge, Massachusetts 02139, USA}
\begin{abstract}
We describe a quantum algorithm for preparing states that encode solutions of non-homogeneous linear partial differential equations. The algorithm is a continuous-variable version of matrix inversion: it efficiently inverts differential operators that are polynomials in the variables and their partial derivatives. The output is a quantum state whose wavefunction is proportional to a specific solution of the non-homogeneous differential equation, which can be measured to reveal features of the solution. The algorithm consists of three stages: preparing fixed resource states in ancillary systems, performing Hamiltonian simulation, and measuring the ancilla systems. The algorithm can be carried out using standard methods for gate decompositions, but we improve this in two ways. First, we show that for a wide class of differential operators, it is possible to derive exact decompositions for the gates employed in Hamiltonian simulation. This avoids the need for costly commutator approximations, reducing gate counts by orders of magnitude. Additionally, we employ methods from machine learning to find explicit circuits that prepare the required resource states. We conclude by studying an example application of the algorithm: solving Poisson's equation in electrostatics.
\end{abstract}
\maketitle

\textit{Introduction---} Following the discovery of quantum algorithms for factoring, database search, and universal simulation of quantum systems \cite{shor1994algorithms,grover1996fast,lloyd1996universal}, decades of work have led to the uncovering of numerous quantum algorithms capable of outperforming existing classical methods. Examples include quantum algorithms for algebraic problems such as Pell's equation and the Jones polynomial \cite{hallgren2007polynomial, freedman2002simulation, aharonov2009polynomial, childs2010quantum}, semi-definite programming \cite{brandao2017quantum, brandao2017exponential}, machine learning \cite{lloyd2014quantum, rebentrost2014quantum, biamonte2017quantum}, and ordinary differential equations \cite{berry2007efficient, berry2015hamiltonian, berry2017exponential, berry2010quantum, berry2014high, berry2017quantum, xin2018quantum, leyton2008quantum,cao2013quantum, montanaro2016finite}. Quantum computers also excel at solving linear systems of equations \cite{harrow2009quantum, wiebe2012quantum, clader2013preconditioned, childs2017quantum, wossnig2018quantum}. Here, given an $N\times N$ sparse matrix $A$ and a vector $\mathbf{b}=(b_1,\ldots,b_N)$, the goal is to find a vector $\mathbf{x}=(x_1,\ldots,x_N)$ satisfying the equation $A\mathbf{x}=\mathbf{b}$. Quantum algorithms for this problem take as input the quantum state $\ket{\mathbf{b}}=\sum_{i=1}^N b_i\ket{i}$ and efficiently perform matrix inversion to prepare the state $\ket{\mathbf{x}}=A^{-1}\ket{\mathbf{b}}$ encoding the solution of the linear system of equations. 

We study a continuous version of this problem where the inputs are a function $f(\mathbf{x})$ over $\mathbb{R}^N$ and a differential operator $A$. In its most general form, $A$ is expressed as a function of the variables and their partial derivatives: $A=A(x_1,\ldots,x_N,\frac{\partial}{\partial x_1},\ldots,\frac{\partial}{\partial x_N})$. The task is to find a function $\psi(\mathbf{x})$ satisfying the linear partial differential equation $A\psi(\mathbf{x})=f(\mathbf{x})$, which is said to be non-homogeneous whenever $f(\mathbf{x})\neq 0 $. In direct analogy to the case of a linear system of equations, a specific solution to the non-homogeneous problem can be found by obtaining the inverse operator $A^{-1}$ and computing the function $\psi(\mathbf{x})=A^{-1}f(\mathbf{x})$.

In this work, we present a quantum algorithm for finding solutions to non-homogeneous linear partial differential equations. More specifically, we show how to solve equations of the form $A\psi(\mathbf{x})=f(\mathbf{x})$, where $A$ is a polynomial in the variables and their partial derivatives. We describe the algorithm in the continuous-variable (CV) model of quantum computing \cite{lloyd1999quantum,braunstein2005quantum}, but the algorithm can be implemented in any model for universal quantum computing. Similarly to quantum algorithms for linear systems of equations, the algorithm takes as input a state $\ket{f}$ encoding the non-homogeneous function and outputs a state $\ket{\psi}$ whose wavefunction is proportional to a specific solution of the partial differential equation. In this sense, the algorithm is a continuous-variable version of the quantum algorithm for linear systems of equations. For differential equations of fixed order, the runtime is polynomial in the dimension -- an exponential improvement over the best known classical techniques for solving partial differential equations. 

The algorithm consists of three stages: preparing fixed resource states in ancillary systems, performing Hamiltonian simulation, and measuring the ancilla systems. Although the algorithm can be carried out using standard methods for gate decompositions, we improve on this in two ways. First, for several cases of interest, we introduce exact decomposition formulas that circumvent the use of commutator approximations for Hamiltonian simulation, leading to shorter circuits by orders of magnitude. Additionally, based on recent results on state preparation using quantum neural networks \cite{killoran2018continuous, arrazola2018machine}, we show how short-depth quantum circuits can be directly optimized to prepare required resource states with high fidelity  Finally, we validate the performance of the algorithm through numerical simulations by studying an example application: solving Poisson's equation in electrostatics.

\textit{Quantum algorithm---} The algorithm takes as inputs (i) a classical description of a linear differential operator $A$ and (ii) a quantum state $\ket{f}$ of $N$ registers with wavefunction $\braket{\mathbf{x}|f}=f(\mathbf{x})$, where $\ket{\mathbf{x}}=\ket{x_1}\ldots \ket{x_N}$. For definiteness, we consider the registers to be modes of the quantized electromagnetic field with associated position $\hat{X}$ and momentum $\hat{P}$ quadrature operators. This choice is for convenience: the algorithm can in principle be carried out in any physical model of quantum computing.

The quadrature operators $\hat{X}_k$ and $\hat{P}_k$ acting on mode $k$ can be defined in terms of their action on an arbitrary state: $\hat{X_k}\int dx^n \psi(\bx) \ket{\bx}=\int dx^N x_k\psi(\bx) \ket{\bx}$ and $
\hat{P}_k\int dx^N \psi(\bx) \ket{\bx}=-\frac{i}{2}\int dx^N \frac{\partial}{\partial x_k}\psi(\bx) \ket{\bx}$ for all $k=1,\ldots, N$, where we have set $\hbar=1/2$. Note that the action of the momentum operator is equivalent to differentiation with respect to position. More generally, a linear differential operator $A$ can be equivalently cast as an operator $\hat{A}$ on the Hilbert space of an $N$-mode quantum system. The operator $\hat{A}$ is then a polynomial of the position and momentum operators. We focus on Hermitian operators, in which case $\hat{A}$ can be viewed as a Hamiltonian for the $N$-mode system. 

We follow the Fourier decomposition technique of Ref.~\cite{childs2017quantum}. Let $g(x)$ be an odd function satisfying $\int_0^\infty g(x)dx=1$. It holds that 
$a^{-1} = \int_0^\infty g(ax)dx$ for $a\neq 0$. Choosing $g(x)=xe^{-x^2/2}$ and writing $g(x)$ in terms of its Fourier transform $g(x)=\frac{i}{\sqrt{2\pi}}\int_{-\infty}^{\infty}dy\, ye^{-y^2/2}e^{ixy}$ we have
\begin{align}
a^{-1}&=\frac{i}{\sqrt{2\pi}}\int_{-\infty}^\infty dx\, \Theta(x) \int_{-\infty}^\infty dy\, ye^{-y^2/2}e^{-iaxy},\label{Eq: 1/x}
\end{align}
where $\Theta(x)$ is the Heaviside step function. Let $\{\ket{a}\}$ be the eigenbasis of $\hat{A}$ with corresponding eigenvalues $a\in\mathbb{R}$. Since $\hat{A}^{-1}$ and $e^{-i \hat{A} xy}$ are both diagonal in the basis $\{\ket{a}\}$, Eq. \eqref{Eq: 1/x} implies that $\hat{A}^{-1}$ can be expressed as
\beq
\hat{A}^{-1}=\frac{i}{\sqrt{2\pi}}\int_{-\infty}^\infty dx dy\, \Theta(x) \, ye^{-y^2/2}e^{-i\hat{A}xy}\label{Eq:Ainv_Fourier}.
\eeq
To implement the action of $\hat{A}^{-1}$ on a target state $\ket{f}$, consider the unnormalized two-mode resource state
\begin{align}
\ket{s}\ket{1}:=\int_{-\infty}^\infty dx \,\Theta(x)\ket{x} \int_{-\infty}^\infty \frac{i}{\sqrt{2\pi}}dy\,  ye^{-y^2/2}\ket{y}\label{Eq:resource_states}.
\end{align}

We refer to the state $\ket{s}$ as a step function state. In its current form, $\ket{s}$ is unnormalizable, but as we discuss shortly, this can be remedied by employing a step function of finite length. Additionally, we recognize $ye^{-y^2/2}$ as the unnormalized wavefunction of a single photon and consequently $\ket{1}$ as a single-photon state up to a global phase and normalization.

\begin{center}
\begin{figure}
\includegraphics[width=0.75\columnwidth]{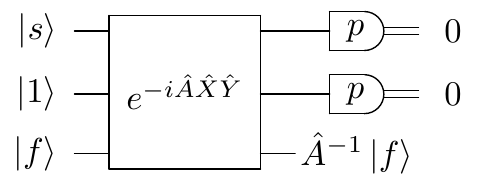}
\caption{Schematic representation of the quantum algorithm. The state $\ket{f}$ is given as input. Two resource states are prepared: a single photon $\ket{1}$ and a step function state $\ket{s}$. A global unitary $e^{-i \hat{A}\hat{X}\hat{Y}}$ is applied to all three systems, which is equivalent to evolution under the Hamiltonian $\hat{A}\hat{X}\hat{Y}$ for unit time. This is followed by a homodyne momentum measurement on the resource modes. Post-selecting on the outcome $p=0$ on both modes yields the desired output state $\hat{A}^{-1}\ket{f}$. }\label{Fig:Algorithm}
\end{figure}
\end{center}

Given an input state $\ket{f}$, the algorithm starts by preparing the resource states of Eq. \eqref{Eq:resource_states}. A global unitary $e^{-i \hat{A}\hat{X}\hat{Y}}$ is subsequently applied to all systems, where $\hat{X}$ and $\hat{Y}$ are respectively the position operators of the two resource modes. This transformation is equivalent to performing evolution under a Hamiltonian $\hat{H}=\hat{A}\hat{X}\hat{Y}$ for unit time. The result is the output state
\beq
\ket{\Psi}=\frac{i}{\sqrt{2\pi}}\int_{-\infty}^\infty dx dy\, \Theta(x) \, ye^{-y^2/2}e^{-i \hat{A}xy}\ket{f}\ket{x}\ket{y}.
\eeq
Performing a momentum homodyne measurement on both resource modes and post-selecting on observing the outcome $p=0$ on both modes, i.e., projecting onto the state $\ket{0_{p_x}}\ket{0_{p_y}}$, yields
\begin{align}
&\left(\id\otimes \ket{0_{p_x}}\bra{0_{p_x}}\otimes\ket{0_{p_y}}\bra{0_{p_y}}\right)\ket{\Psi}=\hat{A}^{-1}\ket{\psi}\ket{0_{p_x}}\ket{0_{p_z}}.
\end{align}
The output is thus the desired state $\ket{\psi}=\hat{A}^{-1}\ket{f}$ with wavefunction $\psi(\bx)=A^{-1}f(\bx)$ up to normalization. The algorithm is depicted in Fig.~\ref{Fig:Algorithm}.

\begin{center}
\begin{figure}[t!]
\includegraphics[width=0.75\columnwidth]{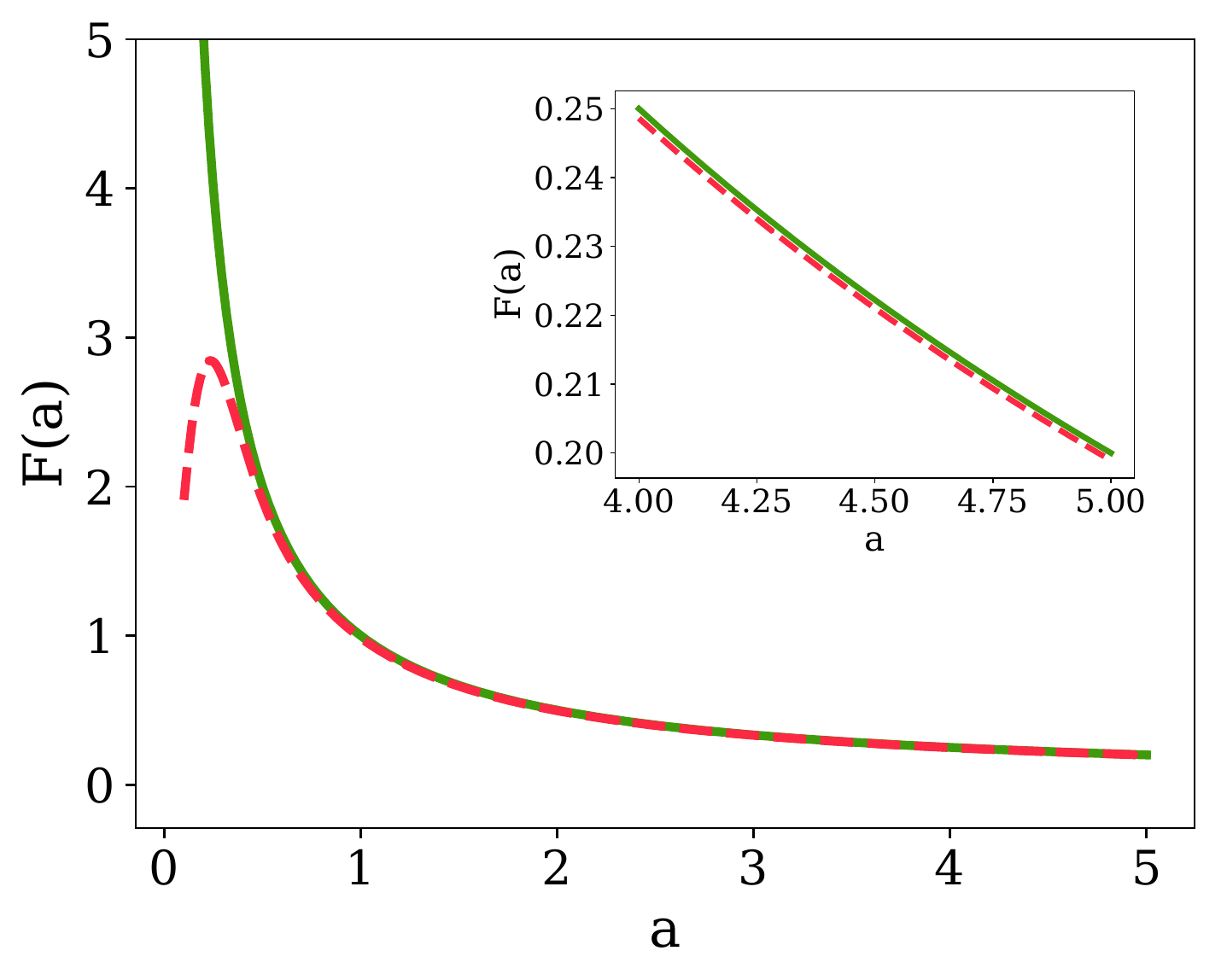}
\caption{The function $a^{-1}$ (solid green) and the approximation $F(a)$ (dashed red) for $L=7$ and $\Delta=0.1$. The inset shows a close-up of the two functions for larger values of $a$. The approximation is excellent except for small values of $a$.  }\label{Fig: Ainv_approx}
\end{figure}
\end{center}

An ideal step function state is unphysical since its wavefunction is not square-integrable. Instead, we consider a step function state of finite width $L$: $\ket{s_L}=\frac{1}{\sqrt{L}}\int_0^{L}dx \ket{x}$. Similarly, homodyne measurements have finite precision, whose effect on the resulting output state must be taken into account. We model this finite-precision measurement as a projection onto finitely squeezed states $\ket{\Delta} = \frac{1}{\pi^{1/4}\sqrt{\Delta}}\int dp\, e^{-p^2/2\Delta^2}\ket{p}$, where $\Delta$ is the measurement precision. The effect of a finite-width step function and finite measurement precision is that, instead of $\hat{A}^{-1}$, the operator applied to the input state $\ket{f}$ is the operator $\hat{A}^{-1}_{\text{approx}}$, which is an approximation to $\hat{A}^{-1}$. The action of this operator can be best understood in terms of its action on an eigenstate $\ket{a}$ of $\hat{A}$. As shown in the Appendix, the result is 
\begin{align}\label{Eq: A_approx_total}
&\hat{A}^{-1}_{\text{approx}}\ket{a}\nonumber=\frac{2a\sqrt{\pi} \Delta(1-e^{-L^2(a^2+\Delta^2+\Delta^4)/2(1+\Delta^2)})}{\sqrt{1+\Delta^2}(a^2+\Delta^2+\Delta^4)}\ket{a}\nonumber\\
&=2\sqrt{\pi} \Delta F(a)\ket{a}.
\end{align}
Here we have implicitly defined the function $F(a)$, which is an approximation to $a^{-1}$. In the limit of $\Delta\rightarrow 0$, the effect of a finite-width step function is to introduce a correction $(1-e^{L^2a^2/2})$ that is only relevant for small values of $a$. The relationship between $F(a)$  and $a^{-1}$ is illustrated in Fig.~\ref{Fig: Ainv_approx} for $L=7$ and $\Delta=0.1$. Despite these being considerable deviations from the ideal case of $L=\infty$ and $\Delta=0$, the approximation is excellent except for small values of $a$. Finally, note that the action of $\hat{A}^{-1}_{\text{approx}}$ on $\ket{f}$ introduces an overall constant factor $2\sqrt{\pi}\Delta$ and therefore the probability of projecting onto the desired output state satisfies $\|\hat{A}^{-1}_{\text{approx}}\ket{f}\|^2=O(\Delta^2)$.

\vspace{-0.5cm}
\textit{Hamiltonian simulation---} The goal of Hamiltonian simulation is to find a quantum circuit that performs the transformation $e^{i \hat{H} t}$ for some Hamiltonian $\hat{H}$ and time $t>0$. The circuit is specified in terms of a universal gate set, which in this work we take to be the set 
\beq\label{Eq: universal_set}
\{e^{i\frac{\pi}{2}(\hat{X}^2+\hat{P}^2)},\, e^{it_1\hat{X}},\, e^{i t_2 \hat{X}^2},\, e^{it_3 \hat{X}^3},\, e^{i\tau \hat{X}_1\otimes\hat{X}_2 }\},
\eeq
where $t_1, t_2, t_3$ and $\tau$ are adjustable real parameters. The Fourier transform gate $\hat{F}:=e^{i\frac{\pi}{2}(\hat{X}^2+\hat{P}^2)}$ has the effect of mapping between the quadrature operators: $\hat{F}^{\dagger}\hat{X}\hat{F}=-\hat{P}$ and $\hat{F}^{\dagger}\hat{P}\hat{F}=\hat{X}$.
The standard approach for performing Hamiltonian simulation is to employ a Trotter-Suzuki decomposition \cite{trotter1959product, suzuki1993general,dhand2014stability, kalajdzievski2018continuous} to express the transformation $e^{it\hat{H}}=e^{it\sum_{j=1}^M \hat{H}_j}$ in terms of the product 
\beq\label{Eq: trotter-suzuki}
e^{it\hat{H}}=\prod_{j=1}^M\left(e^{i\frac{t}{K}\hat{H}_j}\right)^K+O(t^2/K).
\eeq 
Following this, a sequence of commutator approximations are typically employed to decompose each term $e^{i\frac{t}{K}\hat{H}_j}$ into elements from the universal set \cite{sefi2011decompose}. We focus on the case where the Hamiltonian $\hat{A}$ can be expressed as
\beq\label{Eq: quadratic A}
\hat{A} = \lambda\id + \sum_{j=1}^N a_j \hat{X}_j + b_j \hat{P}_j +\alpha_j\hat{X}_j^2+\beta_j\hat{P}_j^2,
\eeq
where $\lambda,a_j, b_j, \alpha_j$, and $\beta_j$ are real constants. This form encompasses a large class of differential operators, including for instance those defining Poisson's equation, the heat equation, and the wave equation. 

In the quantum algorithm, we perform evolution under the Hamiltonian $\hat{A}\hat{X}\hat{Y}$ for unit time. When $\hat{A}$ is of the form of Eq. \eqref{Eq: quadratic A}, after performing a Trotter-Suzuki decomposition as in Eq. \eqref{Eq: trotter-suzuki}, each term in the product will correspond -- up to Fourier transforms exchanging $\hat{X}$ and $\hat{P}$ -- to unitaries of the form $e^{it\hat{X}_j\hat{X}_k}$, $e^{it\hat{X}_j\hat{X}_k\hat{X}_l}$ or $e^{it\hat{X}_j^2\hat{X}_k\hat{X}_l}$, where the subindices denote which mode the operators act on. We now show how exact decompositions can be found for each of these transformations.

First, note that the transformation $e^{it\hat{X}_j\hat{X}_k}$ is already part of the universal set. For the unitary $e^{it\hat{X}_j\hat{X}_k\hat{X}_l}$, it can be shown (see Appendix for details) that the following exact decomposition holds:

\begin{align}\label{Eq: exact_decomp_X1}
&e^{i2t\hat{X}_j\hat{X}_k\hat{X}_l} = e^{i2\hat{P}_j\hat{X}_k} e^{i2\hat{P}_j\hat{X}_l} e^{\frac{it}{3}\hat{X}^{3}_j} e^{-i2\hat{P}_j\hat{X}_l} e^{-i2\hat{P}_j\hat{X}_k}\nonumber\\
 &e^{i2\hat{P}_k\hat{X}_l} e^{\frac{-it}{3}\hat{X}^{3}_k}  e^{-i2\hat{P}_k\hat{X}_l}
e^{i2\hat{P}_l\hat{X}_j} e^{\frac{-it}{3}\hat{X}^{3}_l}  e^{-i2\hat{P}_l\hat{X}_j}\nonumber\\
 &e^{i2\hat{P}_j\hat{X}_k} e^{\frac{-it}{3}\hat{X}^{3}_j}  e^{-i2\hat{P}_j\hat{X}_k}
e^{\frac{it}{3}\hat{X}^{3}_j} e^{\frac{it}{3}\hat{X}^{3}_k} e^{\frac{it}{3}\hat{X}^{3}_l}.
\end{align}
Note that gates of the form $e^{-i2\hat{P}_j\hat{X}_k}$ are equivalent to a controlled-phase gate up to Fourier transforms on the first mode. Finally, as shown in the Appendix, for the gate $e^{it\hat{X}_j^2\hat{X}_k\hat{X}_l}$, it is possible to derive an exact decomposition
\begin{align}\label{Eq: exact_decomp_X1^2}
&e^{i6t\hat{X}_j^{2}\hat{X}_k\hat{X}_l} = e^{i2\hat{P}_k\hat{X}_l} e^{i2\hat{P}_k\hat{X}_j^2} e^{it \hat{X}_k^3}e^{-i2\hat{P}_k\hat{X}_j^2} 
e^{-i2\hat{P}_k\hat{X}_l}\nonumber\\
 &e^{i2\hat{P}_k\hat{X}_j^2} e^{-it \hat{X}_k^3}e^{-i2\hat{P}_k\hat{X}_j^2} e^{i2\hat{P}_k\hat{X}_l} e^{-it \hat{X}_k^3}e^{-i2\hat{P}_k\hat{X}_l}\nonumber \\
 &e^{i2\hat{P}_l\hat{X}_j^2} e^{-it \hat{X}_l^3}e^{-i2\hat{P}_l\hat{X}_j^2}e^{it \hat{X}_k^3}e^{it \hat{X}_j^6}e^{it \hat{X}_l^3}.
\end{align}
Here, unitaries of the form $e^{i2\hat{P}_k\hat{X}_j^2}$ and $e^{it \hat{X}_j^6}$ are not part of the universal set, but exact decompositions can also be derived for them (see Appendix for details). Just which polynomials functions of the quadrature operators are susceptible to such exact decompositions is an interesting open question. The resulting exact decomposition for the gate $e^{it\hat{X}_j^2\hat{X}_k\hat{X}_l}$ contains 873 gates from the universal set. It is important to contrast this with the commutator approximation method \cite{sefi2011decompose}, which requires 28 gates to decompose $e^{it\hat{X}_j^2\hat{X}_k\hat{X}_l}$, but for a precision of $10^{-3}$, it needs about $10^6$ repetitions for a total of roughly $10^7$ gates.  

For any sparse Hamiltonian that is a polynomial of constant degree over the quadrature operators, universal simulation theorems \cite{lloyd1999quantum, braunstein2005quantum} state that $\text{poly}(N)$ time is required to perform Hamiltonian simulation and therefore to run our quantum algorithm for partial differential equations. This is an exponential improvement over classical algorithms for solving PDEs, which scale exponentially with dimension  \cite{thomas2013numerical,werschulz1991computational,ritter1996average}. The runtime of Hamiltonian simulation also scales polynomially on the operator norm $\|\hat{A}\|_{\infty}$~\cite{lloyd1996universal}, so care must be taken to ensure that this norm is well-behaved over the support of the input state $\ket{f}$.

\textit{Resource state preparation---}
We employ results from Refs.~\cite{killoran2018continuous,arrazola2018machine} to find circuits for preparing the single photon and step function resource states required in the algorithm. The strategy is to optimize a quantum neural network which takes a single-mode vacuum state as input and prepares a desired target state as output. A layer $\mathcal{L}$ of the quantum neural network is composed of the sequence of gates~\cite{killoran2018continuous}:  $\mathcal{L}:= K(\kappa) D(\alpha) R(\phi_2) S(r, \theta) R(\phi_1)$ where $R(\theta)$ is a rotation gate, $D(\alpha)$ is a displacement gate, $S(r)$ is a squeezing gate, and $K(\kappa)$ is a Kerr gate. The rotation, squeezing, and displacement gates are Gaussian and can be straightforwardly decomposed in terms of the universal set of Eq. \eqref{Eq: universal_set}. The Kerr gate can be decomposed using results from \cite{sefi2011decompose}. 

\begin{center}
\begin{figure}[t!]
\includegraphics[width=0.5\columnwidth]{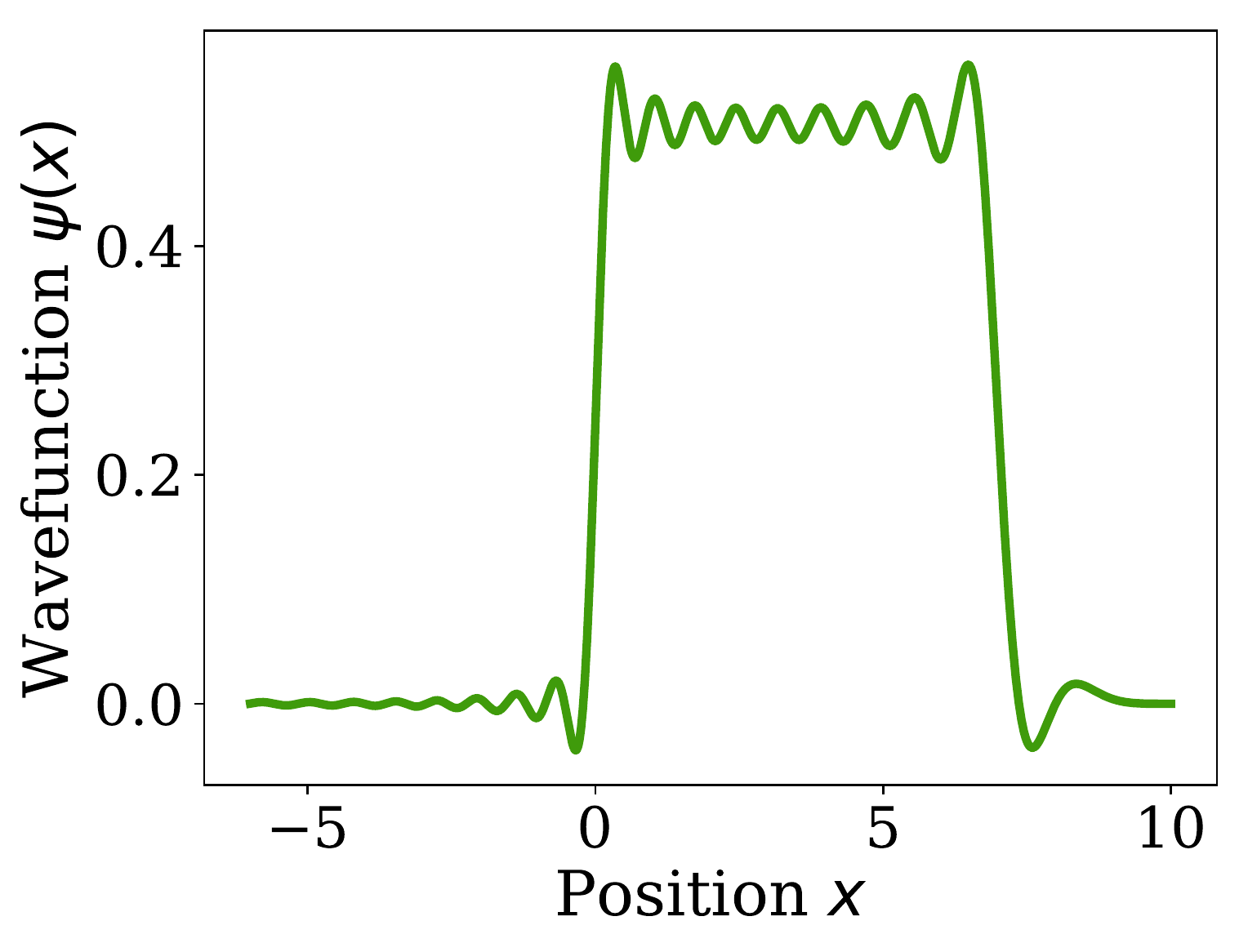}\includegraphics[width=0.5\columnwidth]{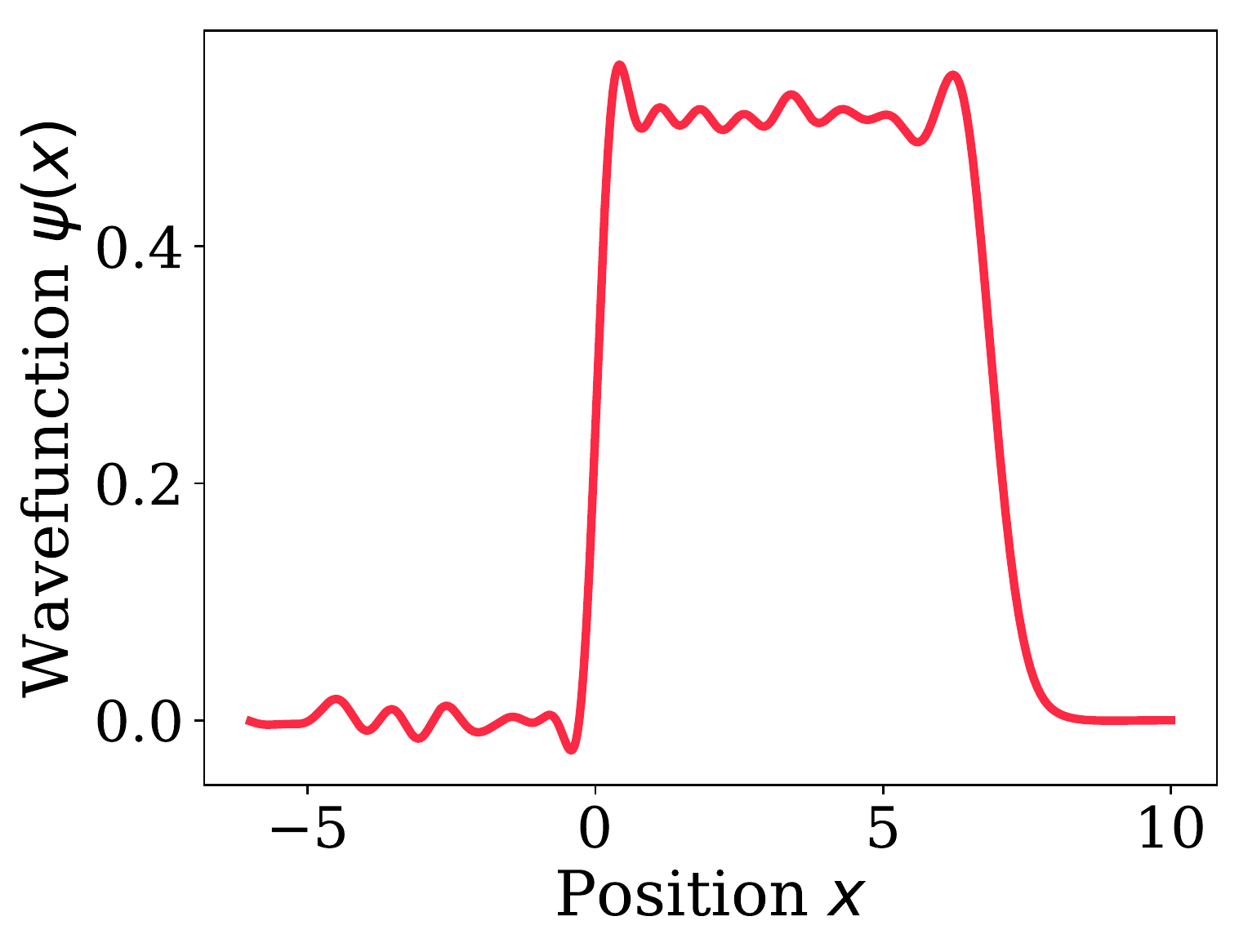}
\caption{(Top) Wavefunctions of the target step function state with cutoff $d=41$ and width $L=7$. (Bottom) The state prepared by the quantum neural network, with fidelity of $99.36\%$ to the target state. The network consists of 30 layers for a total of 150 gates. }\label{Fig: Wavefunctions}
\end{figure}
\end{center}
\vspace{-0.8cm}

We perform optimization of the gate parameters by employing the TensorFlow \cite{abadi2016tensorflow} backend of the Strawberry Fields software platform for photonic quantum computing \cite{killoran2018strawberry}. This approach has been pursued in Ref.~\cite{arrazola2018machine}, where it was shown that a single photon state can be prepared using a quantum neural network of eight layers, i.e., 40 gates, with fidelity $99.998\%$. For the target step function state, we consider the truncated state $\ket{s_L}=\frac{1}{\sqrt{L}}\int_0^Ldx\ket{x}=\sum_{n=0}^\infty c_{n,L}\ket{n}$, where $\ket{n}$ is the Fock state of $n$ photons and $c_{n,L}=\braket{n|s_L}$. For numerical simulations, we introduce a cutoff dimension $d$, yielding the truncated state $\ket{s_{d, L}}=\sum_{n=0}^d c_{n,L} \ket{n}$. As an example, we set a width of $L=7$ and a cutoff $d=41$, fixing a quantum neural network with 30 layers (150 gates) to prepare this state. The result is a network that can prepare a state with $99.36\%$ fidelity to the target state $\ket{s_{41,7}}$. This is shown in Fig.~\ref{Fig: Wavefunctions}, where we plot the wavefunction of both states.
\begin{center}
\begin{figure}[t!]
\includegraphics[width=0.5\columnwidth]{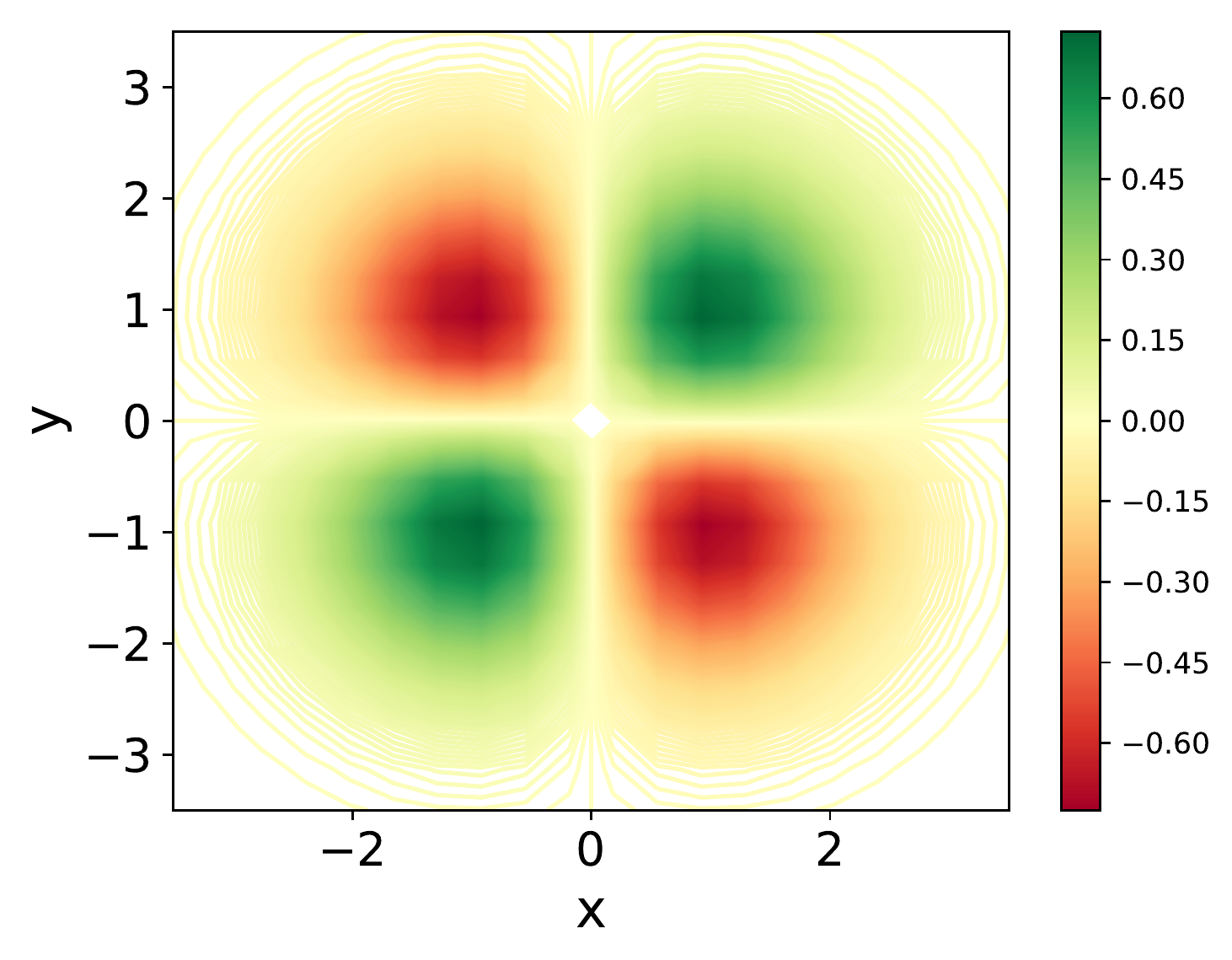}\includegraphics[width=0.5\columnwidth]{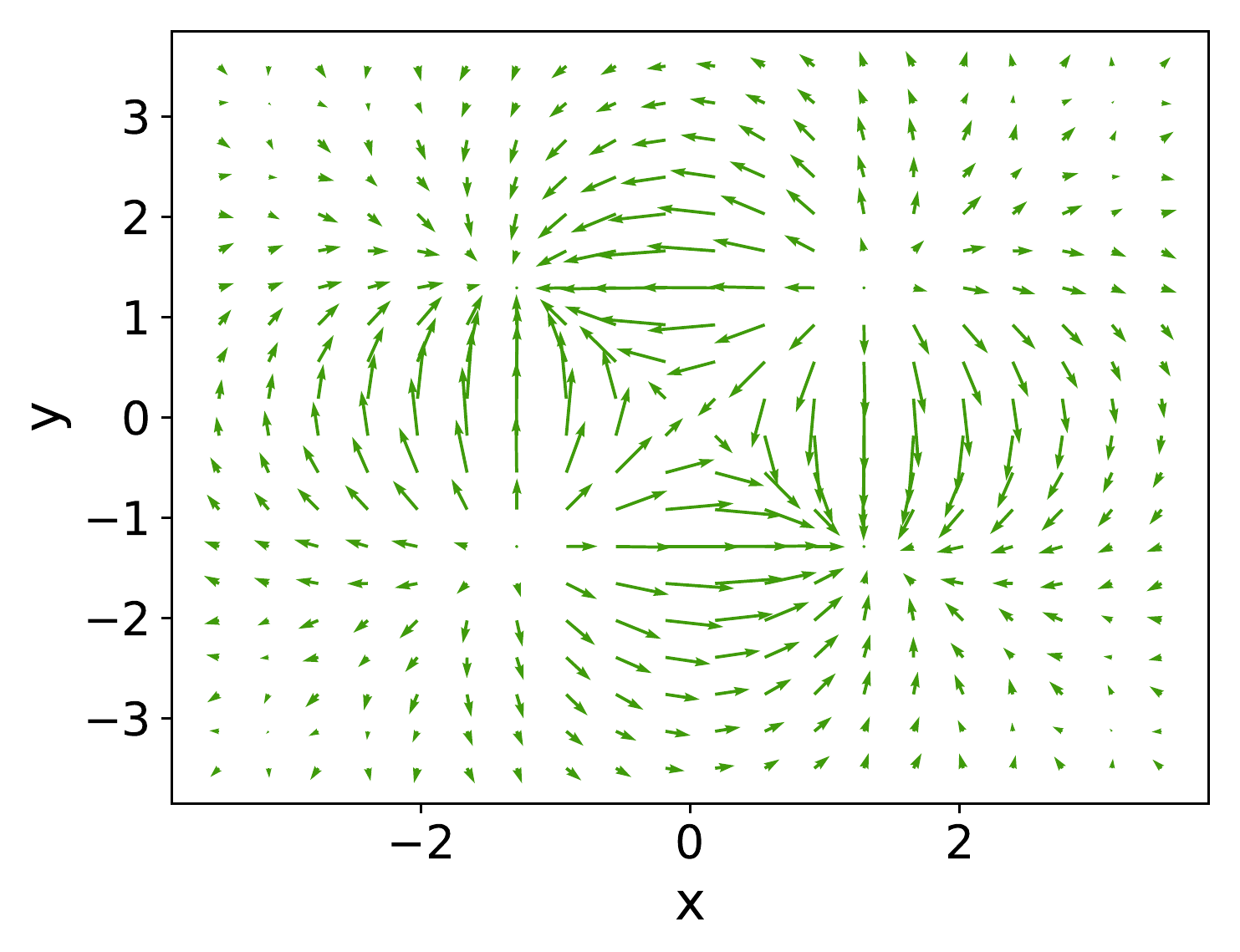}
\caption{(Top) Charge distribution $\rho(x,y)=xy e^{-\frac{x^2}{2}}e^{-\frac{y^2}{2}}$. The top-right and bottom-left quadrants are regions of positive charge while the remaining quadrants are negatively charged. (Bottom) Electric field lines reconstructed from the output state of the quantum algorithm. There are regions of zero electric field in each quadrant that arise from interfering contributions of the charge clouds surrounding these points. The electric field is also zero at the origin, as expected from the symmetry of the charge distribution. }\label{Fig: Poisson}
\end{figure}
\end{center} 

\textit{Example: Poisson's equation---} Poisson's equation is the non-homogeneous partial differential equation 
\beq
\nabla^2\psi(\mathbf{x})=\sum_{i=1}^n\frac{\partial^2 \psi(\mathbf{x})}{\partial x_i^2}=f(\mathbf{x}),
\eeq
which has applications across several areas of physics and engineering. Here we consider its relevance to electrostatics, where it establishes a relationship between a charge distribution $\rho(\mathbf{x})$ and the electric potential $\phi(\mathbf{x})$, namely $
\nabla^2 \phi(\mathbf{x}) = -\frac{\rho(\mathbf{x})}{\varepsilon}$, where $\varepsilon$ is the permittivity of the medium, whose value we fix to $\varepsilon=1$. To apply our quantum algorithm to this problem, note that under the convention $\hbar=1/2$, it holds that $\frac{\partial^2}{\partial x_i^2}=-4\hat{P}^2$ and thus we can set $\hat{A}=-4\sum_{i=1}^n \hat{P}_i^2$. We consider a two-dimensional problem where the charge distribution is given by $\rho(x,y)=xy e^{-\frac{x^2}{2}}e^{-\frac{y^2}{2}}$,
as shown in Fig.~\ref{Fig: Poisson}. This charge distribution is equivalent, up to normalization, to the wavefunction of the two-mode input state $\ket{f}=\ket{1}\ket{1}$ consisting of a single photon in each mode. We compute the output state by constructing the operator $\hat{A}^{-1}_{\text{approx}}$ as in Eq.~\eqref{Eq: A_approx_total} with $L=7$ and $\Delta=0.1$, then applying it to the input state $\ket{f}$. The wavefunction of the output state is proportional to the electric potential $\phi(x,y)$, which can be used to compute the electric field $\vec{E}(x,y)=-\nabla\phi(x,y)$. To illustrate the validity of the solution, the charge distribution and the electric field are shown in Fig.~\ref{Fig: Poisson}, showing the capability of the algorithm to reproduce the desired solution. In a physical implementation of the algorithm, repeated quadrature measurements of the output state would reveal regions of large electrostatic potential.

\textit{Conclusion---} We have presented a quantum algorithm for preparing quantum states that encode the solution to non-homogeneous linear partial differential equations. The algorithm is a continuous-variable version of the quantum algorithm for linear systems of equations. For differential operators of fixed degree, the runtime is polynomial in the dimension $N$. This is an exponential improvement compared to classical algorithms that compute the full solution of partial differential equations. However, there are important differences between this quantum algorithm and classical approaches: the quantum algorithm assumes that the input state $\ket{f}$ encoding the non-homogeneous term of the equation can be efficiently prepared and the output is not an explicit specification of the solution, but instead a state whose wavefunction is proportional to the solution. It is crucial to identify applications where input states can be efficiently prepared and where sampling from the output state -- for example to compute expectation values -- is enough for the task at hand. 

Finally, the quantum algorithm provides a specific solution to the non-homogeneous equation, but to solve general boundary problems it is necessary to also incorporate solutions to the homogeneous equation. Typically, solving the homogeneous equation is less challenging, so in principle the quantum algorithm can be combined with classical methods for solving homogeneous equations to provide full solutions to boundary problems. Future work can focus on finding fully quantum algorithms for boundary problems as well as extending the techniques here presented to more general differential operators.

\acknowledgements

We thank A. Ignjatovic, N. Killoran, T. Bromley and N. Quesada for helpful discussions. S. Lloyd was funded by AFOSR under a MURI on Optimal Quantum Measurements and State Verification, by IARPA under the QEO program, by ARO, and by NSF. 

\appendix

\section{Quantum algorithm}\label{Sec: QAlgorithm}

In the ideal case, the result of the algorithm is the output state
\beq
\ket{\Psi}=\frac{i}{\sqrt{2\pi}}\int_{-\infty}^\infty dx dy\, \Theta(x) \, ye^{-y^2/2}e^{-i \hat{A}xy}\ket{f}\ket{x}\ket{y}.
\eeq
Performing a momentum homodyne measurement on both resource modes and post-selecting on observing the outcome $p=0$ on both modes, i.e., projecting onto the state $\ket{0_{p_x}}\ket{0_{p_y}}$, yields
\begin{align}
&\left(\id\otimes \ket{0_{p_x}}\bra{0_{p_y}}\otimes\ket{0_{p_x}}\bra{0_{p_y}}\right)\ket{\Psi}\nonumber\\
&=\left(\frac{i}{\sqrt{2\pi}}\int_{-\infty}^\infty dx dy\,  \Theta(x) \, ye^{-y^2/2}e^{-i \hat{A}xy}\ket{f}\right)\ket{0_{p_x}}\ket{0_{p_y}}\nonumber\\
&=\left(\hat{A}^{-1}\ket{f}\right)\ket{0_{p_x}}\ket{0_{p_z}}\\
&=\ket{\psi}\ket{0_{p_x}}\ket{0_{p_z}},
\end{align}
where we have used the relation $\braket{0_{p_x}|x}=\braket{0_{p_y}|y}=1$. 

\subsection{Step function state}
An ideal step function state is not square-integrable. Instead, we consider a step function state of finite width $L$ given by
\beq\label{Eq: finite_step_state}
\ket{s_L}=\frac{1}{\sqrt{L}}\int_0^{L}dx \ket{x}.
\eeq 
The result of employing this state in the algorithm is an output state
\begin{align}
\ket{\psi}=\frac{i}{\sqrt{2\pi}}\int_0^Ldx\int_{-\infty}^\infty dy\, ye^{-y^2/2}e^{-i \hat{A}xy}\ket{f}.
\end{align}
In this case, instead of the ideal inverse operator $\hat{A}^{-1}$, the operator being applied to $\ket{f}$ is a truncated Fourier decomposition of $\hat{A}^{-1}$. Here and henceforth we use $\hat{A}^{-1}_{\text{approx}}$ to denote any approximation to the ideal inverse operator.

The effect of a finite width is best understood by considering the action of $\hat{A}^{-1}_{\text{approx}}$ on an eigenstate $\ket{a}$:
\begin{align}\label{Eq: finite_stepfn_approx}
\hat{A}^{-1}_{\text{approx}}\ket{a}=&\frac{i}{\sqrt{2\pi}}\int_0^Ldx\int_{-\infty}^\infty dy\, ye^{-y^2/2}e^{-i \hat{A}xy}\ket{a}\nonumber\\
=&\frac{i}{\sqrt{2\pi}}\int_0^Ldx\int_{-\infty}^\infty dy\, ye^{-y^2/2}e^{-i axy}\ket{a}\nonumber\\
=&\left(\frac{1}{a}-\frac{e^{-a^2L^2}}{a}\right) \ket{a}.
\end{align}

The effect of a truncated step function is an exponentially small correction from the ideal result $a^{-1}\ket{a}$. The correction is only significant for small eigenvalues such that $a\lesssim 1/L$, i.e., the value of $L$ determines the smallest eigenvalue $a$ for which the approximation is adequate.

Even a step function of finite width is an idealization since it is not continuous at either $x=0$ or $x=L$: any physical wavefunction will exhibit a smooth transition around these points. The effect of this finite rise time can be modeled by approximating the ideal step function in terms of a continuous function. Here we consider the error function $\frac{1}{2}(1+\text{erf}[k x])$ as an approximate step function, where the approximation improves with larger $k>0$. We then have
\begin{align}
&\hat{A}^{-1}_{\text{approx}}\ket{a}\nonumber\\
&= \frac{i}{\sqrt{8\pi}}\int_0^\infty dx(1+\text{erf}[k x])\int_{-\infty}^\infty dy\, ye^{-y^2/2}e^{-i \hat{A}xy}\ket{a}\nonumber\\
&=\frac{1}{a}\left(\frac{\sqrt{2}}{\sqrt{2+\frac{a^2}{k^2}}}\right)\ket{a}=\frac{1}{a}\left[1-\frac{a^2}{2k^2}+O\left(\frac{a^4}{k^4}\right)\right]\ket{a}.
\end{align}
This induces another correction from the ideal scenario, but in this case the effect is significant only for large eigenvalues $a$ such that $a\gtrsim k$. Thus, approximations to an ideal step function state lead to deviations that are relevant only for very small or very large eigenvalues.

\subsection{Finite measurement precision}

Homodyne measurements have finite precision, whose effect on the resulting output state must be taken into account. We model this finite-precision measurement as a projection onto finitely squeezed states, as opposed to momentum eigenstates which are infinitely squeezed. The state corresponding to the $p=0$ outcome of a homodyne measurement with precision $\Delta$ is $\ket{\Delta} = \frac{1}{\pi^{1/4}\sqrt{\Delta}}\int dp\, e^{-p^2/2\Delta^2}\ket{p}$. The resulting output state is given by

\begin{align}
&\left(\id\otimes\ket{\Delta}\bra{\Delta}\otimes\ket{\Delta}\bra{\Delta}\right)\ket{\Psi}\nonumber\\
=&\frac{i}{\sqrt{2\pi}}\int_0^{\infty} dx\int_{-\infty}^\infty dy\, ye^{-y^2/2}e^{-i \hat{A}xy}g(x,y)\ket{f}\ket{\Delta}\ket{\Delta}\nonumber\\
=&\left(\hat{A}^{-1}_{\text{approx}}\ket{f}\right)\ket{\Delta}\ket{\Delta}\label{Eq: finite_precision_state}
\end{align}
where
\begin{align}
g(x,y,\Delta)&=\frac{1}{\sqrt{\pi}\Delta}\int_{-\infty}^\infty dp\,dq\,e^{-p^2/2\Delta^2-ipx}e^{-q^2/2\Delta^2-iqy}\nonumber\\
&=e^{-(x^2+y^2)\Delta^2/2}.
\end{align}
As before, the approximation of the inverse operator is best expressed in terms of its action on the eigenstate $\ket{a}$:
\begin{align}
&\hat{A}^{-1}_{\text{approx}}\ket{a}\nonumber\\
=& \frac{i}{\sqrt{2\pi}}\int_0^{\infty} dx\int_{-\infty}^\infty dy\, ye^{-y^2/2}e^{-i \hat{A}xy}g(x,y,\Delta)\ket{a}\nonumber\\
&=\frac{2a\sqrt{\pi} \Delta}{\sqrt{1+\Delta^2}(a^2+\Delta^2+\Delta^4)}\ket{a}\nonumber\\
&= \left[\frac{2\sqrt{\pi} \Delta}{a}+O\left(\frac{\Delta^3}{a^3}\right)\right]\ket{a}.
\end{align}
Up to normalization, this state is equal to the desired state $a^{-1}\ket{a}$ except for a correction $O(\Delta^3/a^3)$ that is only relevant when $a\lesssim \Delta$.

Comparing to Eq. \eqref{Eq: finite_stepfn_approx}, the dominant error for small values of $a$ arises from the finite width of the step function state, whose effect is exponential in $a$ for $a\lesssim 1/L$. Thus, the algorithm can tolerate relatively large values of the measurement precision $\Delta$ without significant consequences. 

By expressing the state $\ket{f}$ in terms of the eigenbasis $\{\ket{a}\}$ of $\hat{A}$, we note that the action of $\hat{A}^{-1}_{\text{approx}}$ on $\ket{f}$ introduces an overall constant factor $2\sqrt{\pi}\Delta$ and therefore the probability of projecting onto the desired output state of Eq. \eqref{Eq: finite_precision_state} satisfies $\|\hat{A}^{-1}_{\text{approx}}\ket{f}\|^2=O(\Delta^2)$. There is thus a tradeoff between the probability of observing the desired outcome and the resulting perturbation of the output state. As discussed before, the correction from the finite measurement precision is less significant than the one due to a finite width of the step function. This makes it possible to select relatively large values of the precision $\Delta$, leading to a higher probability of observing the desired outcome.

Combining the effects of a finite-width step function and a finite-precision measurement leads to an approximation
\begin{align}\label{Eq: A_approx_total_appendix}
&\hat{A}^{-1}_{\text{approx}}\ket{a}\nonumber\\
&= \frac{i}{\sqrt{2\pi}}\int_0^{L} dx\int_{-\infty}^\infty dy\, ye^{-y^2/2}e^{-i \hat{A}xy}g(x,y,\Delta)\ket{a}\nonumber\\
&=\frac{2a\sqrt{\pi} \Delta(1-e^{L^2(a^2+\Delta^2+\Delta^4)/2(1+\Delta^2)})}{\sqrt{1+\Delta^2}(a^2+\Delta^2+\Delta^4)}\ket{a}\nonumber\\
&=2\sqrt{\pi} \Delta F(a)\ket{a},
\end{align}
where we have implicitly defined the function $F(a)$, which is an approximation to $a^{-1}$. 

\section{Exact Decompositions}
\label{appendixdecomps}

We use the convention $[\hat{X},\hat{P}] = i/2$ and begin with the decomposition below:
\begin{widetext}
\begin{align} \label{DecompHHL}
e^{i2\delta\hat{X}_{j}\hat{X}_{k}\hat{X}_{l}} = \: &e^{i2\hat{P}_{j}\hat{X}_{k}} e^{i2\hat{P}_{j}\hat{X}_{l}} e^{\frac{i\delta}{3}\hat{X}^{3}_{j}} e^{-i2\hat{P}_{j}\hat{X}_{l}} e^{-i2\hat{P}_{j}\hat{X}_{k}}
e^{i2\hat{P}_{k}\hat{X}_{l}} e^{\frac{-i\delta}{3}\hat{X}^{3}_{k}}  e^{-i2\hat{P}_{k}\hat{X}_{l}} \nonumber \\
 &e^{i2\hat{P}_{l}\hat{X}_{j}} e^{\frac{-i\delta}{3}\hat{X}^{3}_{l}}  e^{-i2\hat{P}_{l}\hat{X}_{j}}
e^{i2\hat{P}_{j}\hat{X}_{k}} e^{\frac{-i\delta}{3}\hat{X}^{3}_{j}}  e^{-i2\hat{P}_{j}\hat{X}_{k}}
e^{\frac{i\delta}{3}\hat{X}^{3}_{j}} e^{\frac{i\delta}{3}\hat{X}^{3}_{k}} e^{\frac{i\delta}{3}\hat{X}^{3}_{l}}.
\end{align}
\end{widetext}
This equation can be best understood by looking at the right-hand and building each term in sequence. To begin, note that the first four cubic operators in the decomposition are surrounded by operators of the form $e^{i2\hat{P}_{j}\hat{X}_{k}}$ which can be expanded with unitary conjugation:
\begin{equation}
e^{i2\hat{P}_{j}\hat{X}_{k}} e^{i\delta\hat{X}^{3}_{j}} e^{-i2\hat{P}_{j}\hat{X}_{k}} =  e^{i\delta\left(\hat{X}_{j} + \hat{X}_{k}\right)^3}.
\end{equation}
The first one in mode $j$ is translated by the $k$ and $l$ modes, leading to an exponent $\left(\hat{X}_j + \hat{X}_k + \hat{X}_l \right)^3$. Similarly, the other three cubic gates lead to the exponents $\left(\hat{X}_k + \hat{X}_l \right)^3$, $\left(\hat{X}_l + \hat{X}_j \right)^3$, and $\left(\hat{X}_j + \hat{X}_k \right)^3$. Expanding these polynomials gives a series of operators which can be simplified to give the cubic gate on the left hand side.

The second decomposition, from Eq.~(\ref{Eq: exact_decomp_X1^2}) is:
\begin{widetext}
\begin{align} \label{DecompHHL2}
e^{i6\delta\hat{X}_{j}^{2}\hat{X}_{k}\hat{X}_{l}} = \: &e^{i2\hat{P}_{k}\hat{X}_{l}} e^{i2\hat{P}_{k}\hat{X}_{j}^2} e^{i\delta \hat{X}_{k}^{3}}e^{-i2\hat{P}_{k}\hat{X}_{j}^2} 
e^{-i2\hat{P}_{k}\hat{X}_{l}}e^{i2\hat{P}_{k}\hat{X}_{j}^2} e^{-i\delta \hat{X}_{k}^{3}}e^{-i2\hat{P}_{k}\hat{X}_{j}^2} e^{i2\hat{P}_{k}\hat{X}_{l}} e^{-i\delta \hat{X}_{k}^{3}}e^{-i2\hat{P}_{k}\hat{X}_{l}} \nonumber \\
 &e^{i2\hat{P}_{l}\hat{X}_{j}^2} e^{-i\delta \hat{X}_{l}^{3}}e^{-i2\hat{P}_{l}\hat{X}_{j}^2}e^{i\delta \hat{X}_{k}^{3}}e^{i\delta \hat{X}_{j}^{6}}e^{i\delta \hat{X}_{l}^{3}}.
\end{align}
\end{widetext}
For this decomposition we follow the same procedure as above but now use operations $ e^{i2\hat{P}_{k}\hat{X}_{j}^2}$ and $e^{i\delta \hat{X}_{j}^{6}}$ which are not in the universal set. For these, we require the decompositions 
\begin{align} \label{HHL3}
e^{i3\alpha^2 k \hat{P}_{k}\hat{X}_{j}^2 } = \: &e^{ik\hat{P}_{k}^3}e^{-i\alpha \hat{X}_{j} \hat{X}_{k}}e^{-ik\hat{P}_{k}^3}e^{-i2\alpha \hat{X}_{j} \hat{X}_{k}} e^{ik\hat{P}_{k}^3} \nonumber \\
 &e^{i\alpha \hat{X}_{j} \hat{X}_{k}}e^{-ik\hat{P}_{k}^3} e^{i2\alpha \hat{X}_{j} \hat{X}_{k}}e^{i\alpha^3 k \frac{3}{4}\hat{X}_{j}^3},
\end{align}
and
\begin{equation}
e^{i\delta \hat{X}_{j}^{6}} =  e^{i2\hat{P}_{k}\hat{X}_{j}^3} e^{i\delta \hat{X}_{k}^{2}}e^{-i2\hat{P}_{k}\hat{X}_{j}^3} e^{-i\delta \hat{X}_{k}^{2}} e^{-i2\delta\hat{X}_{k}\hat{X}_{j}^3}.
\end{equation}
This final equation also requires a decomposition for the terms of the form $e^{i2\hat{P}_{k}\hat{X}_{j}^3}$  (up to Fourier transform in the second mode). This can be achieved via the expression
\begin{align}
e^{2i\alpha^2 \hat{P}_{k}\hat{X}_{j}^3 } = \: &e^{-i\alpha \hat{X}_{j}^2 \hat{P}_{k}^2} e^{-2i\alpha \hat{X}_{j} \hat{X}_{k}} e^{i\alpha \hat{X}_{j}^2 \hat{P}_{k}^2} \nonumber \\
 &e^{2i\alpha \hat{X}_{j} \hat{X}_{k}} e^{-2i\alpha^3 \hat{X}_{j}^4},
\end{align}
that requires the further decompositions
\begin{align}
e^{i\alpha \hat{X}^{2}_{j}\hat{X}^{2}_{k}} = \: &e^{i2\hat{P}_{j}\hat{X}_{k}}e^{i\frac{\alpha}{12}\hat{X}_{j}^{4}}e^{-i4\hat{P}_{j}\hat{X}_{k}}e^{i\frac{\alpha}{12}\hat{X}_{j}^{4}}e^{i2\hat{P}_{j}\hat{X}_{k}} \nonumber \\
 &e^{-i\frac{\alpha}{6}\hat{X}_{j}^{4}}e^{-i\frac{\alpha}{6}\hat{X}_{k}^{4}}
\end{align}
as well as
\begin{equation}
e^{i\alpha\hat{X}_{k}^4} = e^{2i\hat{P}_{j}\hat{X}_{k}^2} e^{i\alpha\hat{X}_{j}^2} e^{-2i\hat{P}_{j}\hat{X}_{k}^2} e^{-i\alpha\hat{X}_{j}^2}  e^{-2i\alpha\hat{X}_{j}\hat{X}_{k}^2}.
\end{equation}

The total gate count for Eq.~(\ref{DecompHHL2}) in terms of universal gates is 873, but it is exact. If we wish to express all operations in terms of $\hat{X}$ operators only, we can use unitary conjugation with the Fourier transform gate. This will bring the total number of gates from the universal set to 1,749. On the other hand, the standard commutator approximation method requires about 28 gates, but for a precision of $10^{-3}$ will need $10^6$ repetitions for a total of roughly $2.8\times 10^7$ gates to decompose the original operation.

\section{Resource state preparation}\label{Sec: State_prep}
For the step function state, we consider the finite width state $\ket{s_L}$ of Eq.~\eqref{Eq: finite_step_state} as the target state, which can be expressed in terms of the Fock basis as $\ket{s_L}=\sum_{n=0}^\infty c_{n,L} \ket{n}$ for appropriate coefficients $\{c_{n, L}\}$. For numerical simulations, it is convenient to express quantum states in the Fock basis, which requires a cutoff dimension $d$ to be introduced. In this case the target state to be prepared is
\begin{align}
\ket{s_{d,L}} &= \frac{1}{\mathcal{N}}\sum_{n=0}^d c_{n,L} \ket{n}\nonumber\\
&= \frac{1}{\mathcal{N}}\sum_{n=0}^d c_{n,L}\int dx \psi_n(x) \ket{x},
\end{align}
where $\mathcal{N}=\sum_{n=0}^d |c_{n,L}|^2$ is a normalization constant and $\psi_n(x)$ is the wavefunction of the $n$-photon Fock state.

\begin{center}
\begin{figure}[t!]
\includegraphics[width=0.75\columnwidth]{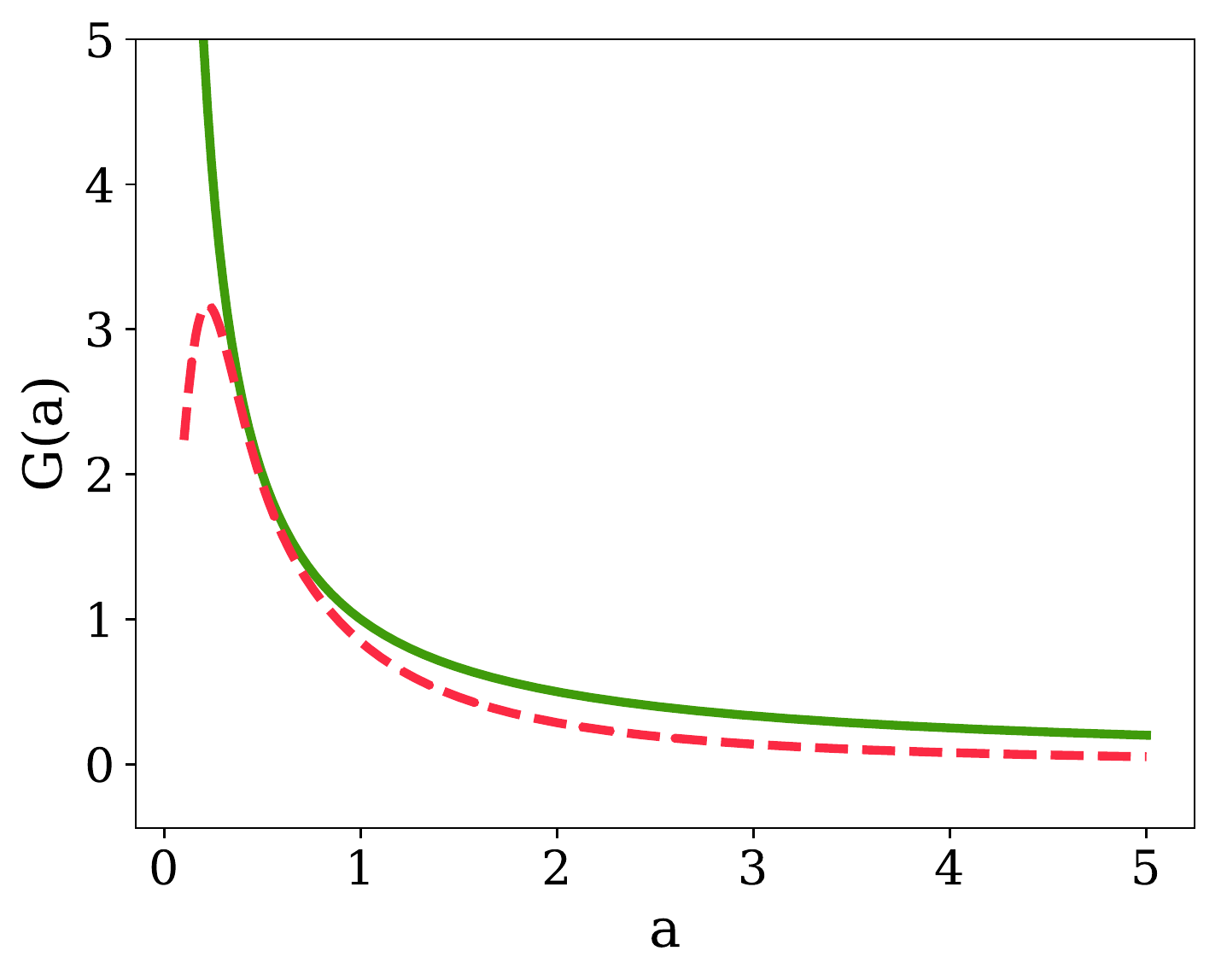}
\caption{The function $a^{-1}$ (solid green) and the approximation $G(a)$ (dashed red) arising from the use of the step function state prepared by the quantum neural network. }\label{Fig: Ainv_wf_approx}
\end{figure}
\end{center}

In principle, a good approximation to the ideal step function state can be obtained by selecting a sufficiently large cutoff dimension, but this leads to an expensive overhead in the simulation of the quantum circuits. As an example, we consider the step function state of width $L=7$ and cutoff $d=41$ and fix a quantum neural network with 30 layers (150 gates) to prepare this state. The result is a network that can prepare a state with $99.36\%$ fidelity to the target state $\ket{s_{41,7}}$. 

As in section \ref{Sec: QAlgorithm}, we evaluate the resulting approximation to the inverse operator $\hat{A}^{-1}$ by computing the function
\beq
G(a)= \frac{i}{\sqrt{2\pi}}\int_{-\infty}^\infty dx \sum_{n=0}^d \gamma_n \psi_n(x) \int_{-\infty}^{\infty}dy\, ye^{-y^2/2}e^{-iaxy},
\eeq
where $\Psi_p(x)=\sum_{n=0}^d \gamma_n\psi_n(x)$ is the wavefunction of the state $\ket{s_p}=\sum_{n=0}^d \gamma_n\ket{n}$ prepared by the network. As shown in Fig.~\ref{Fig: Ainv_wf_approx}, $G(a)$ is also an approximation to the ideal case $a^{-1}$. However, here the approximation deviates more significantly from the ideal case for large values of $a$. As discussed previously, this can be understood from the fact that the wavefunction of the output state is itself an approximation of an ideal step function state of finite width. 
\begin{center}
\begin{figure}[t!]
\includegraphics[width=0.8\columnwidth]{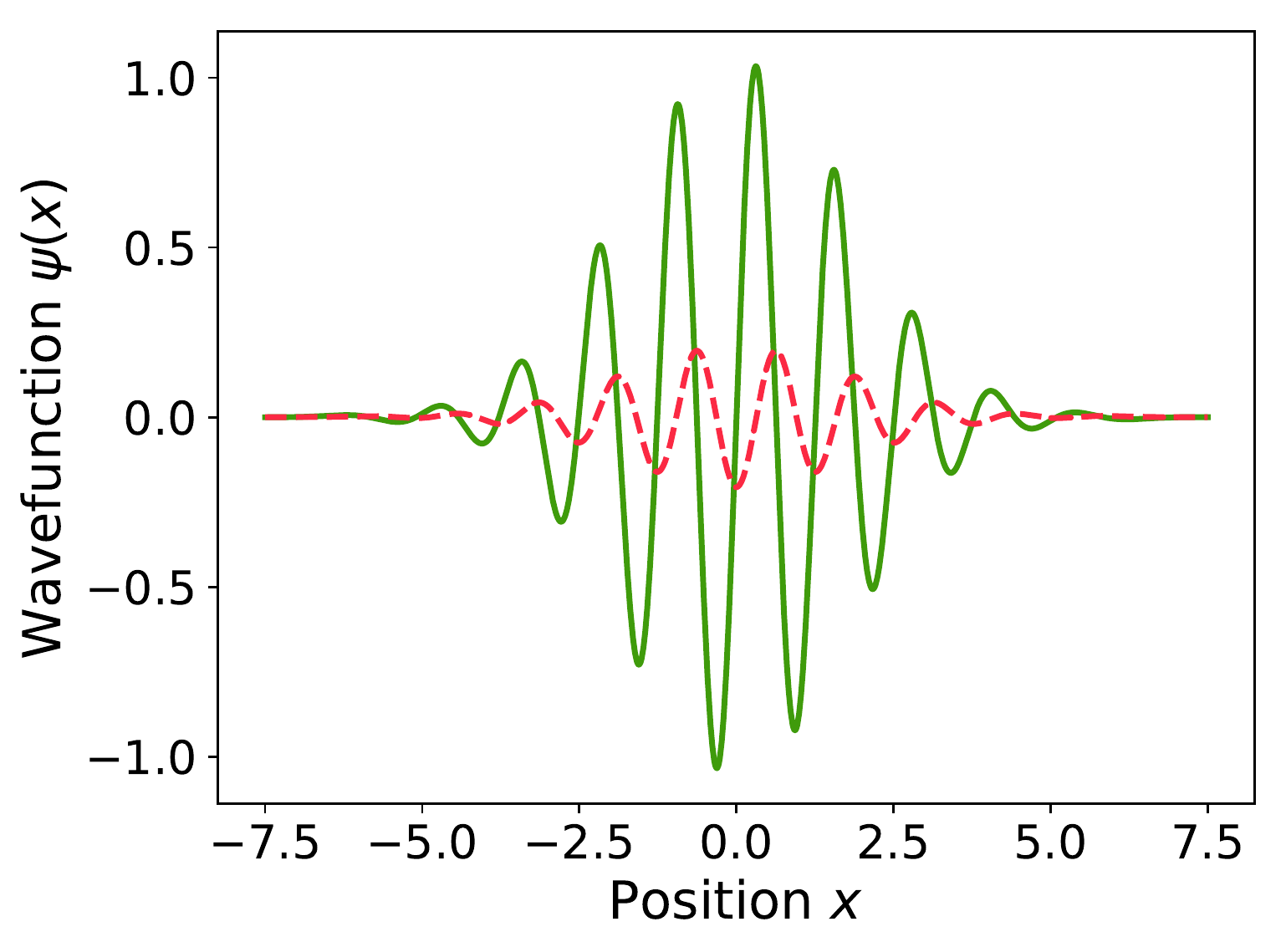}
\includegraphics[width=0.8\columnwidth]{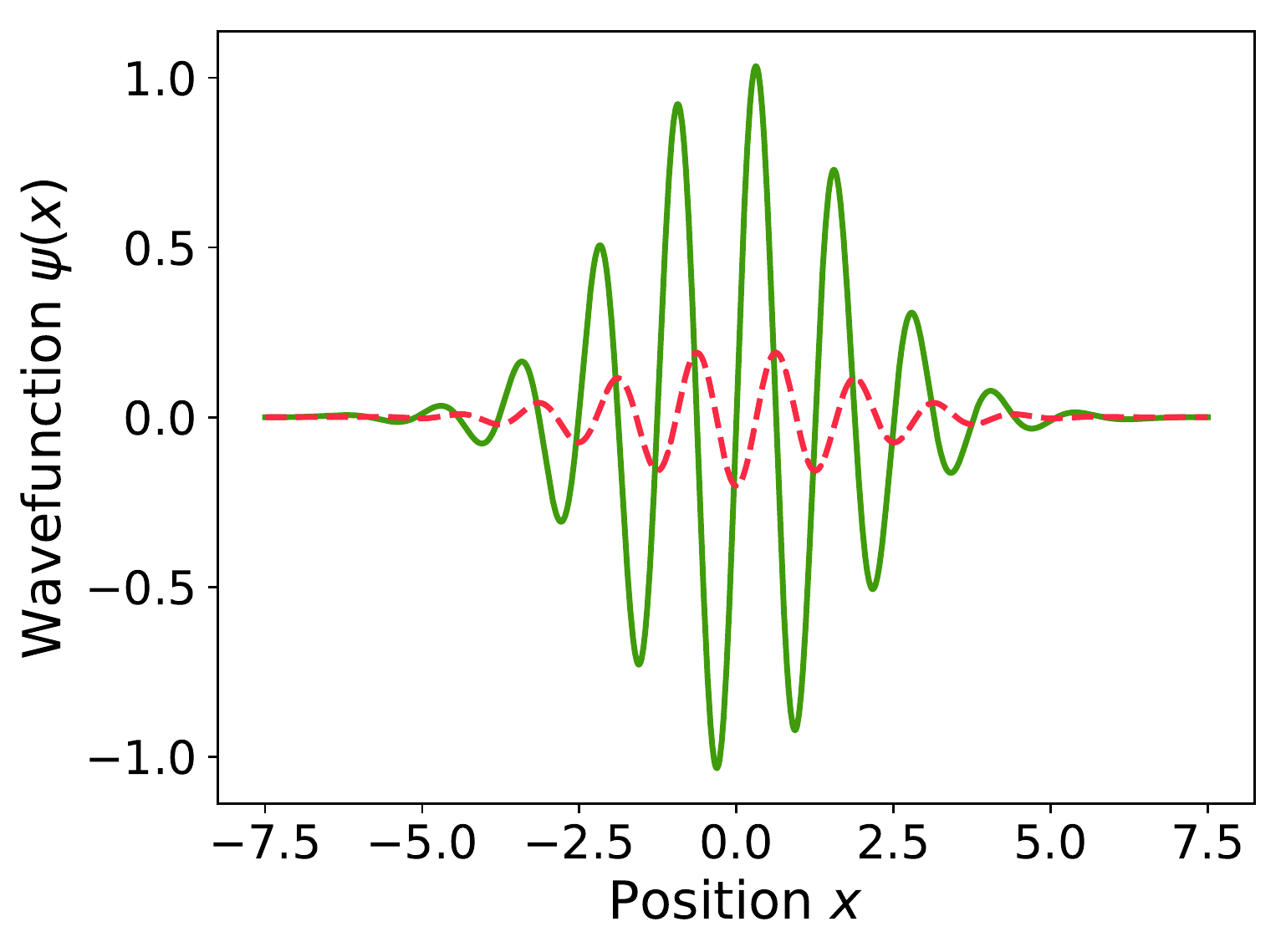}
\caption{(Top) Function $f(x)=\sin(\omega x)e^{-x^2/(2\sigma^2)}$ (solid green) with $\omega=5$ and $\sigma=1.8$. The wavefunction of the output state of the algorithm is also shown (dashed red). (Bottom) Function $f(x)=\sin(\omega x)e^{-x^2/(2\sigma^2)}$ with $\omega=5$ (solid green) and its integral (dashed red), which can be computed analytically. The wavefunction of the output state of the algorithm is almost identical to integral, as desired. In both cases we consider an approximate inverse operator  $\hat{A}^{-1}_{\text{approx}}$ with parameters $L=7$ and $\Delta=0.1$, demonstrating that the effect of a finite-width step function state and finite measurement precision do not significantly affect the correctness of the algorithm. }\label{Fig: Integration}
\end{figure}
\end{center}

\section{Example -- One-dimensional integration}
The simplest non-homogeneous differential equation is the one-dimensional equation $A\psi(x):=\frac{d \psi(x)}{dt}=f(x)$. A solution to the equation is 
\beq
A^{-1} f(x)=\int dx\, f(x),
\eeq
i.e., the solution is the indefinite integral of $f(x)$. To apply the quantum algorithm to this problem we set $\hat{A}=\hat{P}$, in which case the output of the algorithm is the state $\ket{\psi}=\hat{P}^{-1}\ket{f}$ whose wavefunction is proportional to
\beq
\psi(x)=2i\int dx\, f(x).
\eeq
This is equal to the desired solution up to a global phase. In short, the quantum algorithm performs one-dimensional integration. To calculate the output state of the algorithm, we numerically compute the operator 
\beq
\hat{A}^{-1}_{\text{approx}}=\frac{i}{\sqrt{2\pi}}\int_0^{L} dx\int_{-\infty}^\infty dy\, ye^{-y^2/2}e^{-i \hat{A}xy}g(x,y, \Delta)
\eeq
as in Eq.~\eqref{Eq: A_approx_total_appendix} and consequently calculate $\hat{A}^{-1}_{\text{approx}}\ket{f}$.

This computation is performed by expressing $\hat{P}$ in the Fock basis, truncating to a finite photon number, and approximating the integral by a Riemann sum. The operator $\hat{A}^{-1}_{\text{approx}}$ includes the effects of a finite-width step function state and limited precision measurement, but not of approximations to the step function state. We choose $f(x)=\sin(\omega x)e^{-x^2/(2\sigma^2)}$  as the function to integrate. The corresponding input state $\ket{f}=\sum_n c_n \ket{n}$ can be obtained by computing the coefficients $c_n=\int_{-\infty}^{\infty}dx f(x)\psi_n(x)$, where $\psi_n(x)$ is the wavefunction of the Fock state with $n$ photons.

The results are shown in Fig.~\ref{Fig: Integration} where we plot $f(x)=\sin(\omega x)e^{-x^2/(2\sigma^2)}$ and the wavefunction of the output state. The output wavefunction closely reproduces the integral of $f(x)$, even when considering a finite-width step function state with parameter $L=7$ and measurements with finite precision $\Delta=0.1$.

\bibliographystyle{apsrev}
\bibliography{Bibliography}

\end{document}